\documentclass[12pt,preprint]{aastex}
\usepackage{emulateapj5}
\usepackage{graphicx}
\usepackage{longtable}
\usepackage{ulem}
\usepackage{float}
\slugcomment{printed at \today, accepted by ApJ}

\newcommand{\msun}{{M}_{\sun}}

\newcommand{\be}{\begin{equation}}
\newcommand{\ee}{\end{equation}}

\shorttitle{Origin of radio emission in LLAGNs}
\shortauthors{Liu et al.}

\font\mbf = cmmib10 scaled\magstep1
       \font\mbfs = cmmib10 \font\mbfss = cmmib10 scaled 833
\begin{document}

\title{Possible Origin of Radio Emission from Nonthermal Electrons in Hot Accretion Flows for Low-Luminosity Active Galactic Nuclei}

\author{Hu Liu and Qingwen Wu\altaffilmark{*}}
\altaffiltext{1}{School of Physics, Huazhong University of Science and Technology,
 Wuhan 430074, China; \\$^{*}$ Corresponding author, Email: qwwu@hust.edu.cn}

\begin{abstract}
  The two components of radio emission, above and below 86 GHz respectively, from the Galactic center
  source$-$Sgr A* can be naturally explained by the hybrid of thermal and nonthermal electrons
  in hot accretion flows (e.g., radiatively inefficient accretion flow, RIAF, e.g., Yuan et al. 2003).
  We further apply this model to a sample of nearby low-luminosity active galactic nuclei (LLAGNs),
  which are also believed to be powered by the RIAF. We selected the LLAGNs with only compact radio cores
  according to the high-resolution radio observations, and the sources observed with jets
  or jet-like features are excluded. We find that the radio emission of LLAGNs
  is severely underpredicted by pure RIAF model, and can be naturally explained
  by the RIAF model with a hybrid electron population consisting of both thermal and nonthermal particles.
  Our model can roughly reproduce the observed anti-correlation between the mass-corrected radio loudness
  and Eddington ratio for the LLAGNs in our sample.
  We further model the spectral energy distributions of each source in our sample, and
  find that roughly all sources can be well fitted if a small fraction of the steady state electron
  energy is ejected into the nonthermal electrons. The size of radio emission region of our model
  is around several thousand gravitational radii, which is also roughly consistent with the recent
  high-resolution VLBI observations for some nearby LLAGNs.

\end{abstract}


\section{Introduction}

Active galactic nuclei (AGNs) produce enormous luminosities in extremely compact volumes.
It was found some local galaxies show similarities to those of
bright AGNs, but with relatively weaker broad-line emission and multiwavelength radiative cores
\citep[e.g.,][]{ho97,ho01,fa00,na02,ma07}. These low-luminosity active
galactic nuclei (LLAGNs) are very common in the local universe, which may be the scaled-down
luminosity version of bright AGNs \citep[see][for a recent review]{ho08}. Actually, the AGN may only
spend a small fraction of its lifetime in the highly luminous, QSO-like phase, and much more time
is spent in a weakly accreting state \citep[LLAGNs or normal galaxies, e.g.,][]{hh06}.
Both bright AGNs and LLAGNs are believed to be powered by the matter falling onto a super
massive black hole (BH). The observational properties of LLAGNs are quite different from those
of bright QSOs, which are thought to be driven by different accretion modes.
The optical/UV bumps observed in QSOs can be naturally interpreted as
multi-temperature blackbody emission from a cold, optically thick,
and geometrically thin standard disk \citep[SSD,][]{ss73}. However, most of
LLAGNs lack the evident optical/UV bump as that of bright AGNs. The
possible physical reason is that the standard thin accretion disk is absent in these LLAGNs.
A hot, optically thin, geometrically thick radiatively inefficient accretion flow model has been
developed in the last several decades (RIAF, an `updated' version of the original advection dominated accretion
flow model, ADAF; e.g., Ichimaru 1977;
Narayan \& Yi 1994, 1995; Abramowicz et al. 1995; see Kato et al. 2008;
Narayan \& McClintock 2008; and Yuan \& Narayan 2013 for reviews). The RIAF model can successfully
explain most observational features of nearby LLAGNs (e.g., Quataert et al. 1999; Cao \& Rawlings 2004; Nemmen et al. 2006;
Wu et al. 2007; Wu \& Gu 2008; Gu \& Cao 2009; Yuan et al. 2009a; Yuan et al. 2009b; Xu \& Cao 2009;
Ho 2009; Cao 2010; Yu et al. 2011; Nemmen et al. 2012; see Yuan 2007 and Ho 2008 for recent reviews).

The AGNs are traditionally divided into radio loud (RL) and radio quiet (RQ),
and the origin for this dichotomy is still a matter of debate.
The distinction between the RL and RQ objects is normally based on
a radio-loudness parameter, which was defined as
the ratio of the monochromatic flux density at 5~GHz and optical B
band at 4400~$\rm \AA$ ($R_{\rm o}=F_{\rm 5\ GHz}/F_{\rm B}$).
The loudness $R_{\rm o}=10$ is usually taken as the division between RQ and RL AGNs
\citep[particularly in QSO studies, e.g.,][]{ke94}. \citet{tw03} introduced a new definition of the radio loudness
parameter by comparing the 5 GHz radio luminosity to the 2-10 keV luminosity,
$R_{\rm X}=L_{\rm 5GHz}/L_{\rm 2-10keV}$, and proposed that $\log R_{\rm X}=-4.5$ would be the barrier
separating RL and RQ AGNs, which roughly corresponding to $R_{\rm o}=10$. The use of X-ray
luminosity with respect the optical one should largely avoid extinction problems which normally occur
in the optical band. However, most of the traditionally RQ
LLAGNs become RL according to the criterion $R_{\rm o}=10$ or $\log R_{\rm X}=-4.5$ if using their nuclear
emission at radio and optical/X-ray waveband \citep[e.g.,][]{ho02,pa06}. \citet{pa06} redefined the boundary for the
RL and RQ LLAGNs as $\log R_{\rm X}\sim-2.8$ based on a sample of low-luminosity Seyferts and
Fanaroff-Riley type I radio galaxies (FR Is). In the last decade, there is great progress in estimating
the BH mass in both normal galaxies and AGNs \citep[e.g.,][]{ge00,ka05}.  The measurements of BH mass help us to
further understand the issue of radio loudness. \citet{ho02} found that the
loudness parameter increases with decreasing of the Eddington ratio. \citet{si07} investigated the radio
loudness of a total 199 sources which include broad-line radio
galaxies (BLRGs), RL QSOs, Seyferts, low-ionization nuclear
emission-line region galaxies (LINERs), FR Is, and Palomar-Green QSOs (PGQs), and found that
there are two distinct, approximately parallel tracks on
radio-loudness$-$Eddington-ratio plane. They further proposed a quantitative definition
for the radio-loudness parameter that is dependent on the Eddington ratio (see their equation 5),
where the RL sources include BLRGs, RL QSOs, FR Is, while RQ sources comprise Seyferts, LINERs
and PGQs (hereafter we use this criteria to distinguish the RL/RQ).

The radio emission of RL AGNs is believed to originate in the relativistic jets,
where many large-scale radio jets have been observed directly.
Compared with RL AGNs, the origin of nuclear radio emission from RQ objects is still
not well understood, since that RQ AGNs are much fainter radio sources, and most of them only show a
compact radio core. The radio variability of RQ QSOs confined the radio emission to
$\lesssim$0.1~pc \citep[e.g.,][]{ba05}, and the high-resolution VLBI observations
show that the region of the radio emission in RQ LLAGNs should be $\lesssim 10^{-4}$ to $\lesssim 10^{-2}$
pc \citep[e.g.,][]{hu01,na02,gp09}. Several possibilities have been
proposed to explain the radio emission in these RQ AGNs. The first explanation is that RQ AGNs are
scaled-down version of the RL AGNs where exist a weak, small-scale jet \citep[e.g.,][]{mi93,fa96}. A small fraction
of RQ objects show core-jet or linear structures at pc-scale or sub-pc scale, which do
support the weak-jet scenario \citep[e.g.,][]{hu01,ul05,le06}. The accretion-jet model is frequently used to
model the multiwavelength spectral energy distribution (SED) of AGNs \citep[particularly in LLAGNs, e.g.,][]{yu09,yyh11,ne12}.
However, this scenario cannot answer the bimodality of RL/RQ distribution for AGNs \citep[e.g,][]{ke94,si07}.
The second possibility is that the radio emission of the RQ AGNs come from the hot optically thin plasma
in the disk winds, where the plasma is completely ionized and has a density high enough for bremsstrahlung emission
that make a significant contribution to the observed compact radio core \citep[e.g.,][]{bk06,st11}.
In this model the mass-loss rate in the winds are significant and the observed luminosities of radio
emission from quasars imply that they should accrete at super-Eddington rates. However, most of RQ AGNs
are accreting at sub-Eddington rates, where the radio core emission should not dominantly come
from the disk winds. The third possibility is that the radio emission also come from
the accretion flows (e.g., RIAF or disk corona). \citet{lb08} found that the tight correlation between
the radio and X-ray luminosities in RQ QSOs is similar to that of active stars ($L_{\rm R}\sim 10^{-5}\it L_{\rm X}$).
They, therefore, proposed that both the radio and X-ray emission from the nuclei of RQ AGNs may dominated by the
magnetic reconnection heated corona. \citet{wu05} found that the radio emission of
most nearby LLAGNs is higher than that predicted by radiation from the pure
thermal electrons in the RIAF, and their radio emission should have other origin (e.g., nonthermal electrons in RIAF or jet).
The radio spectrum of Sgr A*, a supermassive BH in our galaxy, consists of two components, which
dominate below and above 86 GHz, respectively. \citet{yu03} found that the component above 86 GHz can
be well explained by the thermal electrons from the RIAF, while low-frequency radio spectrum can be explained
if there exist a small fraction of nonthermal electrons in the RIAF \citep[see also][]{oz00}.

It is natural that both the thermal and nonthermal electrons exist in the hot plasma
(RIAF or disk corona), since that the turbulence, magnetic reconnection, and weak shocks can
accelerate electrons and generate a nonthermal tail at high energies in the distribution function
of thermal electrons. The Sgr A* provide an excellent observational evidence for the existence of
hybrid of thermal and nonthermal electrons in accretion flows (e.g., RIAF), since that no evident radio
jet was observed up to now and, therefore, its radio emission may be dominated by other mechanisms.
It is interesting to note that there is also independent observational evidence for nonthermal electrons
in accretion flows. \citet{mc00} reported a high-energy tail
in the hard state of the X-ray binary, Cygnus X-1, extending from
50 keV to $\sim$5 MeV. The data at MeV energies, collected with the COMPTEL instrument of the Compton
Gamma-Ray Observatory, can be explained by a power-law distribution of nonthermal electrons
in the RIAF/corona \citep[e.g.,][]{ro10}. In this work, we try to explore whether the nonthermal elctrons in
RIAF can explain the radio emission of nearby LLAGNs, which always have only
compact radio cores.

\section{Sample}

\begin{table*}[t]
\footnotesize
\begin{minipage}{144mm}
\caption{The sample of LLAGNs}
\label{sample}
\begin{tabular}{lccccccccl}
\tableline\tableline

Name & $D_{\rm L}$ & $\log M_{\rm BH}$ & $\log L_{\rm 5GHz}$ & Telescope & $\log L_{\rm 2-10 keV}$ & Telescope & $\dot{m}_{\rm out}$ & $\eta$ & Refs. \\
     &     Mpc     &   $\msun$   &  $\rm erg/s$    &      & $\rm erg/s$     &      &           &         &      \\
     (1) & (2) & (3) & (4) & (5) & (6) & (7) & (8) & (9) & (10)\\
\tableline
NGC 1097 & $14.5$ & 8.10 & $35.98$ & VLA & $40.63$ & C & 1.1$\times10^{-3}$ & 0.03\% & 1,2,3 \\
NGC 2787 & $13.0$ & 8.07 & $37.06$ & VLBA & $39.28$ & C & 3.0$\times10^{-4}$ & 10.0\% & 4,4,5 \\
NGC 3147 & $40.9$ & 8.60 & $37.78$ & VLBA & $41.88$ & C & 5.3$\times10^{-3}$ & 0.5\% & 4,4,6 \\
NGC 3169 & $19.7$ & 7.79 & $37.16$ & VLBA & $41.41$ & C & 6.0$\times10^{-3}$ & 3.0\% & 4,4,6 \\
NGC 3226 & $23.4$ & 8.07 & $37.06$ & VLBA & $40.74$ & C & 1.1$\times10^{-3}$ & 3.0\% & 4,4,6 \\
NGC 3227 & $20.6$ & 7.44 & $36.31$ & VLBA & $41.70$ & X & 2.1$\times10^{-2}$ & 0.4\% & 4,4,7 \\
NGC 3414 & $24.9$ & 8.42 & $36.59$ & MERLIN & $39.89$ & C & 4.2$\times10^{-4}$ & 0.3\% & 8,8,5 \\
NGC 3718 & $17.0$ & 7.85 & $36.77$ & VLBA & $40.44$ & C & 1.4$\times10^{-3}$ & 3.0\% & 4,4,9 \\
NGC 3941 & $18.9$ & 7.55 & $35.69$ & VLA & $39.27$ & X & 5.5$\times10^{-4}$ & 0.5\% & 10,10,7 \\
NGC 3998 & $21.6$ & 8.72 & $38.34$ & VLBA & $41.79$ & X & 2.3$\times10^{-3}$ & 3.0\% & 4,4,11 \\
NGC 4138 & $17.0$ & 7.51 & $36.13$ & VLA & $41.48$ & X & 1.5$\times10^{-2}$ & 0.1\% & 10,10,7 \\
NGC 4143 & $17.0$ & 8.25 & $37.15$ & VLBA & $40.04$ & C & 5.0$\times10^{-4}$ & 3.0\% & 4,4,6 \\
NGC 4168 & $16.8$ & 7.98 & $36.63$ & VLBA & $39.32$ & X & 3.8$\times10^{-4}$ & 4.0\% & 4,4,12 \\
NGC 4203 & $9.7$  & 7.77 & $36.70$ & VLBA & $39.70$ & C & 6.0$\times10^{-4}$ & 5.0\% & 4,4,13 \\
NGC 4477 & $16.8$ & 8.00 & $35.46$ & VLA & $39.60$ & X & 4.95$\times10^{-4}$ & 0.04\% & 10,10,7 \\
NGC 4565 & $9.7$  & 7.46 & $36.26$ & VLBA & $39.56$ & C & 7.0$\times10^{-4}$ & 7.0\% & 4,4,6 \\
NGC 4594 & $9.2$  & 8.46 & $37.79$ & VLA & $39.95$ & C & 3.2$\times10^{-4}$ & 15.0\% & 14,15,5 \\
NGC 4639 & $16.8$ & 6.68 & $35.57$ & VLA & $40.18$ & C & 4.5$\times10^{-3}$ & 4.0\% & 10,10,13 \\
NGC 4698 & $16.8$ & 7.41 & $35.64$ & VLA & $38.70$ & C & 3.5$\times10^{-4}$ & 3.0\% & 10,10,5 \\
NGC 4736 & $4.3$  & 6.98 & $35.51$ & VLA & $38.77$ & C & 5.7$\times10^{-4}$ & 3.0\% & 1,4,16 \\
\tableline
\end{tabular}
\tablecomments{
Column 1: galaxy name; Column 2: galaxy distance; Column 3: BH mass; Column 4: monochromatic power at 5 GHz;
Column 5: the telescope for radio data; Column 6: 2-10 keV X-ray luminosity; Column 7: the telescope for X-ray data, where
'C' denote $Chandra$ and 'X' denote $XMM-Newton$; Column 8 and 9: model parameters resulting from the SED fits;
Column 10: references for the distance, radio luminosity and X-ray luminosity respectively. \\ References:
(1) \citet{er10}; (2) \citet{th00}; (3) \citet{ne06}; (4) \citet{na05}; (5) \citet{go09};(6) \citet{tw03}; (7) \citet{ca06};
(8) \citet{fi06}; (9) \citet{sa05}; (10) \citet{hu01}; (11) \citet{pt04}; (12) \citet{pa06}; (13) \citet{ho01};
(14) \citet{ho99}; (15) \citet{hu84}; (16) \citet{er02}.
 }
\end{minipage}
\end{table*}

  To examine whether the radio emission of nearby LLAGNs with only compact cores can be explained by the
nonthermal electrons in the RIAF, we search the literatures for high-resolution radio and X-ray data to ensure
that their emission being from the nuclei of sources. In the radio band, we only selected the
sources that have been observed by high-resolution radio telescopes (e.g., VLA, MERLIN, VLBA and VLBI), which are
mainly chosen from \citet{hu01}, \citet{fi06} and \citet{na05}, where all these works tried to
give a complete radio imaging survey of all nearby LLAGNs given in Palomar spectroscopic survey \citep[][]{ho97}.
For purpose of our study, we exclude the sources that observed with radio
jet or even linear radio structures where the radio emission may mainly originate in the jets/outflows. The 5 GHz
nuclear radio luminosities are shown in Table (1), where two sources observed at other waveband (NGC 1097 at 8.4 GHz
and NGC 4736 at 15 GHz) was converted to 5 GHz by assuming $F_{\nu}\propto\nu^{-0.5}$
as found in \citet{ho01} for LLAGNs. In the X-ray waveband, we
only selected the sources have compact X-ray cores, which are observed by high-resolution telescopes of $Chandra$
and/or $XMM-Newton$. Several sources with the high-resolution optical data from $HST$ are also included in building
the SEDs \citep[][and references therein]{er10}. \citet{ma05} gave lower limits for the intrinsic AGN
optical emission for a sample of LINERs, which were constrained from their optical variabilities. We also include
the lower limits of the intrinsic AGN optical emission for three sources
\citep[NGC 3998, NGC 4594 and NGC 4736,][]{ma07}.
   We estimate the BH mass from the velocity dispersion $\sigma_{*}$
of the host bulges for LLAGNs in our sample using the $M_{\rm BH}-\sigma_{*}$ relation given by \citet{tr02}.
Most of velocity dispersions are selected from the Hyperleda database\footnote{http://leda.univ-lyon1.fr} with one
exception, NGC 1097, from \citet{er10}. We find that the velocity dispersion of most LLAGNs in our sample is consistent
with that reported in \citet{hg09}. Our sample include 20 LLAGNs (see Table 1).

\section{Model}
  In this work, we consider the RIAF model proposed by \citet{yu03}, which
can be considered as an updated version of the original ADAF \citep[e.g.,][]{ny95}, where both outflows
and the possible existence of nonthermal electrons are considered. Here, we briefly summarize the model
as follows.

The more accurate global structure and dynamics of the accretion flow is calculated
numerically to obtain the ion and electron temperature, density at each radius.
In particular, we employ the approach suggested by \citet{man00} for calculating the structure of the RIAF in
general relativistic frame, which allows us to calculate the structure of the accretion flow surrounding
either a spinning or a nonspinning BH. In this work, we calculate the global structure of
the accretion flows surrounding massive Schwarzschild black holes.
In stead of using a constant accretion rate, we assume that the accretion rate is a function
of radius, e.g., $\dot{M}=\dot{M}_{\rm out}(R/R_{\rm out})^{\it s}$ \citep[e.g.,][]{bb99,st99,ig99,st01,lc09,yb10},
where $R_{\rm out}$ is the outer radius of the RIAF and $\dot{M}_{\rm out}$ is the accretion rate at $R_{\rm out}$.
The global structure of the RIAF can be calculated with proper outer boundaries, if the parameters
$\dot{m}$, $\alpha$, $\beta$, and $\delta$ are specified  \citep*[see][for more details]{man00}.
All radiation processes (synchrotron, bremsstrahlung and Compton scattering) are included self-consistently
in the calculations of the RIAF structure.
The parameter $\dot{m}=\dot{M}/\dot{M}_{\rm Edd}$ is
dimensionless accretion rate, where the Eddington accretion rate is
defined as $\dot{M}_{\rm Edd}=1.4\times10^{18}M_{\rm BH}/\msun \rm\
gs^{-1}$. For the viscosity parameter $\alpha$ and magnetic parameter $\beta=P_{\rm g}/P_{\rm tot}$
[defined as the ratio of the gas to the total pressure (sum of gas and magnetic pressure)], we
adopt typical values of $\alpha=0.3$ and $\beta=0.9$, which are widely used in RIAF models.
There is obviously a degeneracy between $\dot{m}_{\rm out}$ and $s$ when the accretion rate at
the innermost region of the RIAF is concerned. We adopt $s=0.3$ in this work, which is well constrained
from the observation of our Galactic center black hole, Sgr A* \citep[][but see also \citet{yu12}
for slightly higher value of $s\sim0.4-0.5$ from simulations]{yu03}. We fix the outer boundary
at $5000R_{\rm g}$ ($R_{\rm g}=GM_{\rm BH}/c^2$ is gravitational radius), which is roughly consistent
with the prediction of the disk-corona evaporation model for the LLAGNs with very low Eddington ratios
 \citep[e.g., $10^{-7}\lesssim L_{\rm 2-10\ keV}/L_{\rm Edd}\lesssim 10^{-4}$,][]{lt09}.
The most poorly constrained parameter is $\delta$, which describes the fraction of the
turbulent dissipation that directly heats the electrons in the flow. \citet[][]{sh07} found
that the parameter $\delta$ may be in range of $\sim0.01-0.3$ based on the simulations,
depending on the model details. In this work, we adopt $\delta=0.1$ because we find that the X-ray
spectra of most LLAGNs in our sample can be well fitted with this fixed $\delta$ value.

Following \citet{yu03}, we assume that the injected energy in the nonthermal electrons is
equal to a fraction $\eta$ of the energy in thermal electrons, where we take $\eta$ to
be independent of the radius.
The thermal electrons are assumed to have relativistic Maxwell-Boltzmann distribution, while the
nonthermal electrons are assumed to be the power-law tail of the thermal electrons, which can be described by
the parameter $p$ [$n(\gamma)\propto \gamma^{-p}$,
where $\gamma$ is the electron Lorentz factor]. The number density of the nonthermal electrons can be obtained
if $\eta$ is given. \citet{oz00} proposed that the
parameters, $p$ and $\eta$, for describing the nonthermal electrons are degenerate in
the radio waveband in our model. Therefore, we adopt a fiducial value of $p=3.0$, and allow the parameter $\eta$
to be free.

After determining the distribution of thermal and nonthermal electrons, we calculate
their radiation. The synchrotron
emission from both thermal and nonthermal electrons are calculated \citep[e.g.,][]{yu03}.
The Comptonization of seed photons from both thermal and nonthermal electrons are considered, which
was calculated by the method proposed by \citet{cb90}. In spectral calculations,
the gravitational redshift effect is considered, while the relativistic optics
near the black hole is neglected. More details about spectrum calculation
can be found in \citet[][and references therein]{yu03}.

In summary, our model have only two free parameters, $\dot{m}_{\rm out}$ and $\eta$ in fitting the
SED of LLAGNs.

\section{Results}

\begin{figure}[H]
\centering
\begin{minipage}{90mm}

\includegraphics[width=90mm]{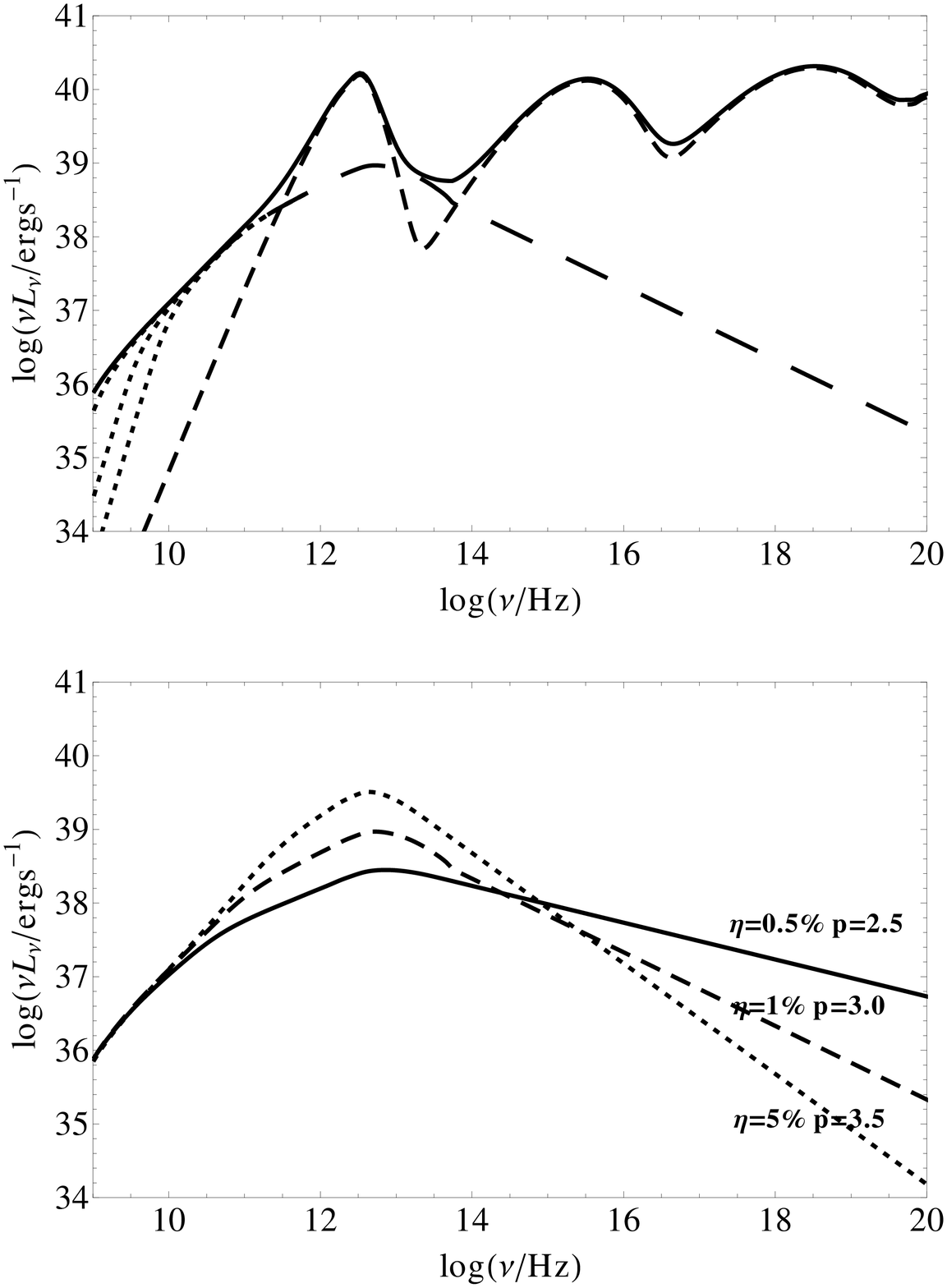}
\end{minipage}
\caption{The top panel show the typical spectrum of the RIAF with both thermal and nonthermal electrons for
$M_{\rm BH}=10^8 \msun$, $\dot{m}_{\rm out}=10^{-3}$, $\eta=1\%$ and $p=3.0$ at given $R_{\rm out}=5000R_{\rm g}$,
where the short-dashed, long-dashed and solid lines are the radiation from the thermal electrons,
nonthermal electrons and sum of them respectively. The dotted lines represent the radiation
of the nonthermal electrons in the RIAF with different outer radius at $3000R_{\rm g}$, $1000R_{\rm g}$,
and $500R_{\rm g}$ (from top to bottom) respectively. The bottom panel show the spectrum from the nonthermal
electrons with $p=2.5$, $\eta=0.5\%$ (solid line); $p=3.0$, $\eta=1\%$ (dashed line), and $p=3.5$, $\eta=5\%$ (dotted
line) respectively, where the other parameters are the same as above.
}
\label{f1}
\end{figure}

  We present the typical spectrum of the RIAF with a population of hybrid electrons in Figure \ref{f1} (top panel),
  where the short-dashed, long-dashed and solid lines represent the emission from the thermal electrons,
  nonthermal electrons and sum of them respectively, where typical parameters $M_{\rm BH}=10^8 \msun$,
  $\dot{m}_{\rm out}=10^{-3}$, and $\eta=1\%$ are adopted for given outer radius $R_{\rm out}=5000 R_{\rm g}$.
  In this model, the X-ray emission mainly come from the Comptonization of seed
  photons that produced by both thermal electrons and nonthermal electrons.
  We find that the low-frequency radio emission is dominated by the self-absorbed synchrotron
  emission from the nonthermal electrons, which is normally 2-3 order of magnitude
  higher than that from thermal electrons in RIAF. Our results are similar for other
  parameters if the typical parameter $\eta\sim$ several percent is adopted as that used
  in Sgr A* \citep[][]{yu03,yu06}. The high-energy X-ray emission dominantly come from the inner region
  of the accretion flows (e.g., within several tens gravitational radii). However,
  we find the region of the radio emission is much larger. The dotted lines in Figure \ref{f1} (top panel) show
  the spectrum from the nonthermal electrons with outer boundary $R_{\rm out}=3000R_{\rm g}$, $1000R_{\rm g}$
  and $500R_{\rm g}$ respectively for above given other parameters. We find that the radiation region of
  the low-frequency radio emission at several GHz can up to several thousands gravitational radii, where the radio
  emission at $\gtrsim3000R_{\rm g}$ is negligible. The bottom panel of Figure \ref{f1} show the spectrum
  of nonthermal electrons with $p=2.5$, $\eta=0.5\%$ (solid line); $p=3.0$, $\eta=1\%$ (dashed line) and
  $p=3.5$, $\eta=5\%$ (dotted line), where the other parameters keep the same as above. It can be found that
  $p$ and $\eta$ are degenerate for radio emission \citep[e.g., $\lesssim 10$ GHz, see also][]{oz00}. Therefore,
  it should be impossible to determine $p$ and $\eta$ from the observational low-frequency radio emission alone.

\begin{figure}[H]
\begin{center}
\begin{minipage}{90mm}
\centering
\includegraphics[width=90mm]{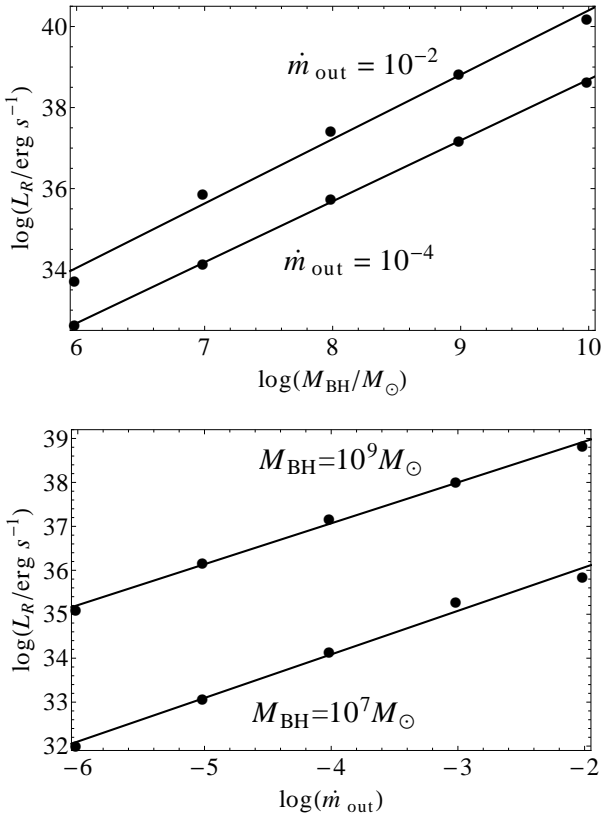}
\end{minipage}
\caption{Top panel show the model prediction on the relation between $L_{\rm 5GHz}$ and
$M_{\rm BH}$ for two given accretion rates $\dot{m}_{\rm out}=10^{-2}$ and $\dot{m}_{\rm out}=10^{-4}$ with $\eta=1\%$,
where the solid points represent $M_{\rm BH}=10^6$, $10^7$, $10^8$, $10^9$, and $10^{10}\msun$ respectively,
and the solid lines are their best fittings. The bottom panel present the relation between $L_{\rm 5GHz}$ and $\dot{m}_{\rm out}$ for two given
BH masses $M_{\rm BH}=10^7$ and $M_{\rm BH}=10^9\msun$, where the solid points represent
$\dot{m}_{\rm out}$=$10^{-6}$, $10^{-5}$, $10^{-4}$, $10^{-3}$, and $10^{-2}$ respectively,
and solid lines are their best fittings.   }
\label{f2}
\end{center}
\end{figure}

\begin{figure}[H]
\begin{center}
\begin{minipage}{90mm}
\centering
\includegraphics[width=90mm]{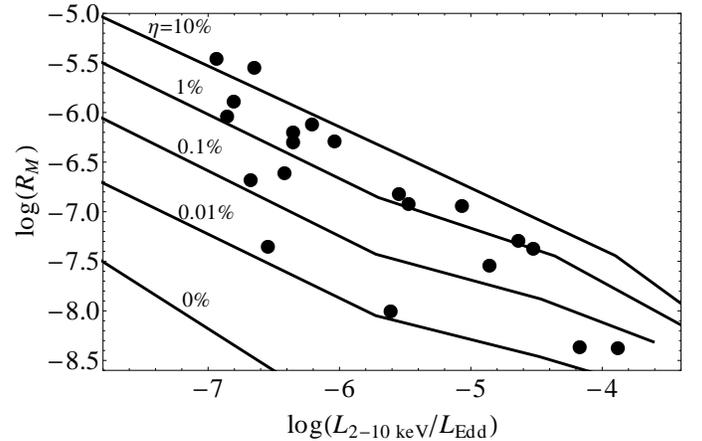}

\end{minipage}
\caption{The relation between the mass-corrected radio loudness, $R_{\rm M}=R_{\rm X}/M_{\rm BH}^{0.4}$,
and the Eddington-scaled X-ray luminosity, $L_{\rm 2-10keV}/L_{\rm Edd}$, where
$R_{\rm X}=L_{\rm 5GHz}/L_{\rm 2-10keV}$. The solid points represent the LLAGNs in
our sample, and the solid lines represent the model prediction with the parameter $\eta$= 10\%, 1\%, 0.1\%, 0.01\%,
 and 0\% (from top to bottom) respectively.}
\label{f3}
\end{center}
\end{figure}

  Our calculations show that the X-ray emission from the RIAF is almost proportional to the BH mass for a given
  $\dot{m}$, and it scales as $\dot{m}^{q}$ with $q\simeq2$ for a given BH mass \citep[see also][]{wu06,mhd03}. In this work,
  we further explore the possible relation between the radio luminosity and BH mass/accretion rate
  for RIAF model with both thermal and nonthermal electrons, where the
  radio emission dominantly originate in the nonthermal electrons.  Figure \ref{f2} (top panel) shows the relation
  between radio luminosity and BH mass for $\dot{m}_{\rm out}=10^{-2}$ and $\dot{m}_{\rm out}=10^{-4}$ with $\eta=1\%$  respectively,
  where we find that the radio emission is roughly proportional to $M_{\rm BH}^{1.4}$ (solid lines).
  The bottom panel of Figure \ref{f2} shows the relation between the radio luminosity and dimensionless accretion rate
  for given BH masses $M_{\rm BH}=10^{7}$ and $10^{9}\msun$ respectively,
  where we find that the radio emission is roughly proportional to $\dot{m}^{\xi}$ with $\xi\simeq0.8$ for both cases.
  To explore the radio emission for a sample of LLAGNs with different BH masses as a whole, we define a mass-corrected
  radio-loudness as $R_{\rm M}=R_{\rm X}/M_{\rm BH}^{0.4}$, according to the dependence of the radio
  and X-ray emission on the BH mass in our model, where $R_{\rm X}=L_{\rm 5GHz}/L_{2-10\rm keV}$.
  We present the  relation between the mass-corrected loudness, $R_{\rm M}$, and the Eddington ratio,
  $L_{\rm 2-10keV}/L_{\rm Edd}$, for all LLAGNs in Figure \ref{f3}, where both quantities roughly
  do not depend on the BH mass. It can be found that there is an anti-correlation between $R_{\rm M}$ and
  $L_{\rm 2-10keV}/L_{\rm Edd}$ for LLAGNs. We further demonstrate the theoretical relation
  between $R_{\rm M}$ and $L_{\rm 2-10keV}/L_{\rm Edd}$ in our model for different values of $\eta$ from 0 to 10\%
  at given $M_{\rm BH}=10^8\msun$ (see Figure \ref{f3}). It can be found that our theoretical model
  also predict that $R_{\rm M}$ is anti-correlate with $L_{\rm 2-10keV}/L_{\rm Edd}$ as
  observational features of LLAGNs. This is because the loudness $R_{\rm M}$ of our model scales
   $R_{\rm M}\propto L_{\rm R}/L_{\rm X}\propto \dot{m}^{-1.2}$, while $L_{\rm X}/L_{\rm Edd}\propto \dot{m}^{2}$.
   We find that the radio emission of most LLAGNs can be explained by our model if
  $\sim 0.01-10$\% of the electron energy is ejected into the nonthermal electrons (see Figure \ref{f3}).
  However, pure RIAF without power-law electrons ($\eta=0$, see Figure \ref{f3}) always
  underpredicts the radio emission by several orders of magnitude compared the observations.

\begin{figure*}
\begin{center}
\begin{minipage}{144mm}
\centering
\includegraphics[width=144mm]{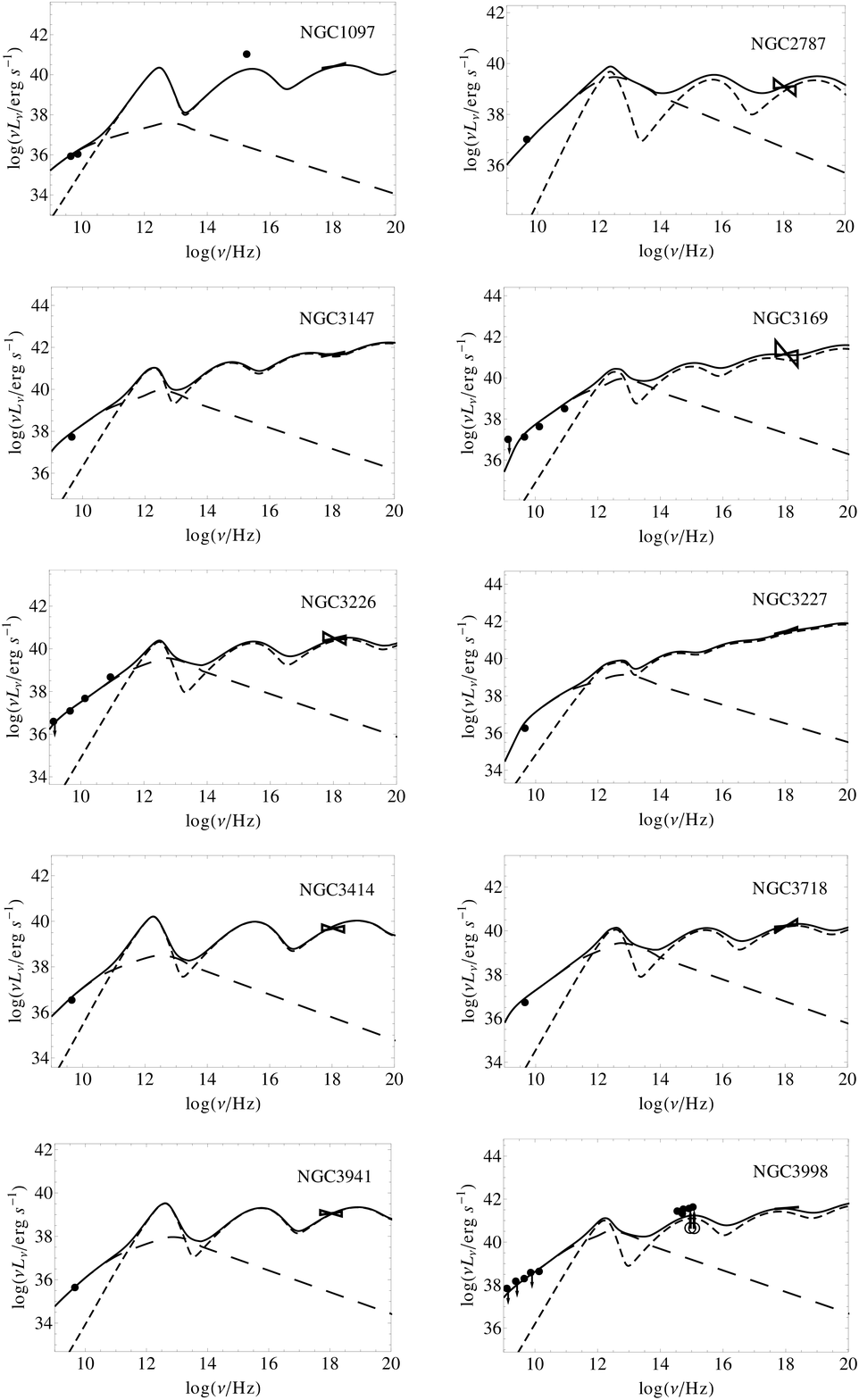}

\end{minipage}
\caption{Models for the SED of LLAGNs. The solid points represent the observed emission, and the empty circles
denote the lower limit for the intrinsic AGN optical emission that derived from the variability of optical emission.
 The short-dashed and long-dashed lines represent the emission
from the thermal and nonthermal electrons of the RIAF respectively, while the solid line show the radiation
from both of these two components.}
\label{f4}
\end{center}
\end{figure*}

\begin{figure*}
\begin{center}
\begin{minipage}{144mm}
\centering
\includegraphics[width=144mm]{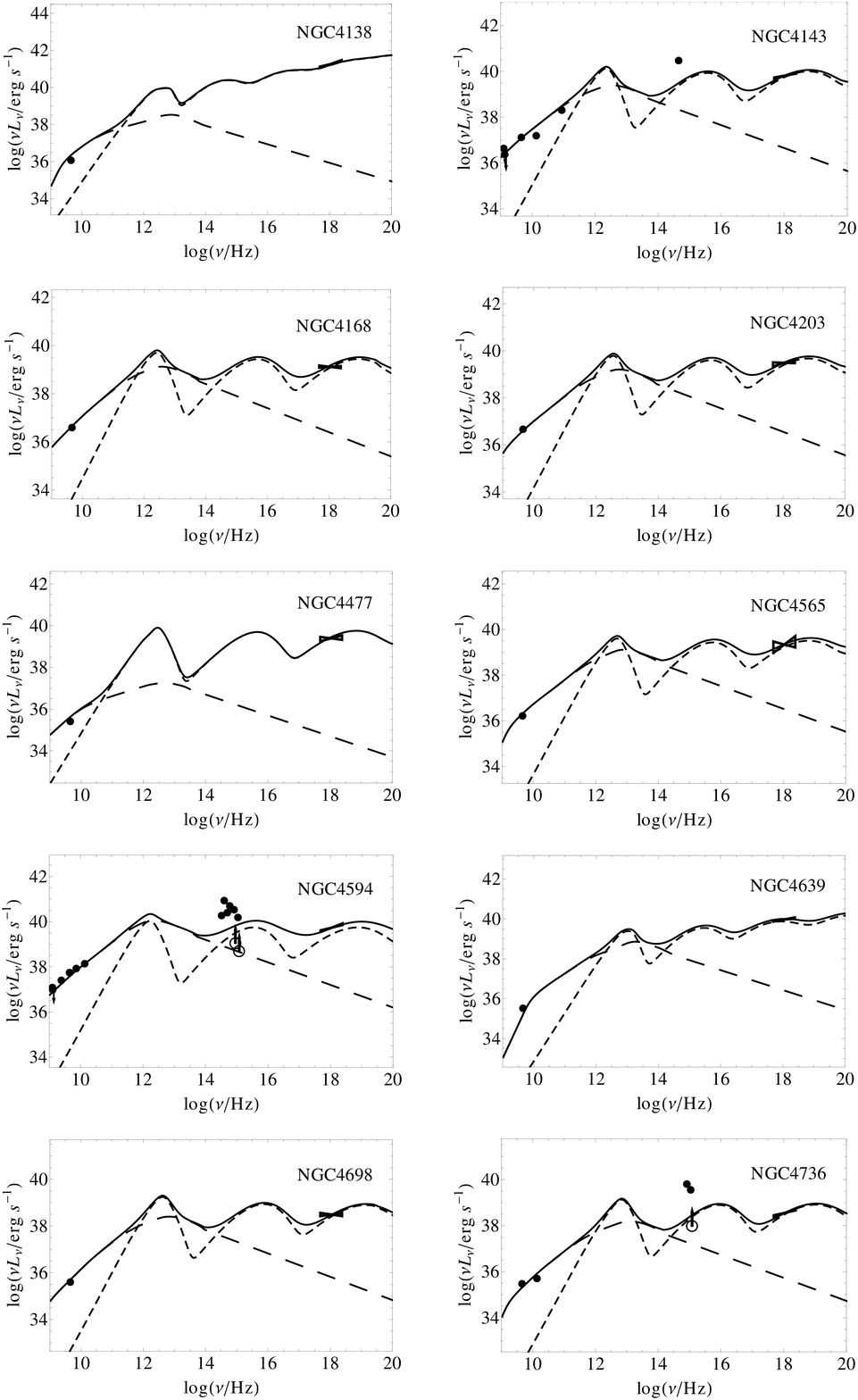}

\end{minipage}
\caption{The same as Figure \ref{f4}, but for the remaining 10 LLAGNs.}
\label{f5}
\end{center}
\end{figure*}

  We further fit the radio to X-ray spectrum of all sources in our sample using the RIAF model
  with hybrid electrons. The results are shown in Figures \ref{f4}-\ref{f5}, and
  the adopted parameters, $\dot{m}_{\rm out}$ and $\eta$, of each source are listed in Table (1). With exception of the
  optical/UV waveband in several sources, the model can successfully fit the SED of these sources
  from the radio to hard X-rays with only two free parameters. The X-ray emission dominantly
  come from the thermal electrons of the RIAF for all sources in our sample, while the observed radio emission is always 2-3
  orders of magnitude higher than that predicted by the purely thermal RIAF model. The radio
  emission can be well described by the synchrotron radiation from the nonthermal electrons in the
  RIAF if a small fraction of the electron thermal energy resides in the power-law tail, where the
  $\eta$ range from 0.03\% to 15\%, with a typical value of $\sim$3\%, for the sources in our present sample (see Table 1).

\section{Summary and Discussion}

  In this work, we explore the hybrid thermal-nonthermal synchrotron emission from the hot accretion flows of the RIAF.
  Such hot accretion flows are expected to exist in sources with low-mass accretion rates (e.g., LLAGNs).
  We calculate the spectrum of the global RIAF model with hybrid electrons, and find that synchrotron emission
  from the nonthermal electrons can be up to 2-3 orders of magnitude greater than that from the purely thermal electrons even
  with only several percent of electron thermal energy injected into the nonthermal electrons.
  This model has explained the multi-wavelength spectrum of Sgr A* very well \citep[][]{yu03}, and we further extend this
  model to other nearby LLAGNs, which have only a compact radio core and have no evidence of the jet.
  Our model can roughly reproduce the observed anticorrelation between the mass-corrected radio
  loudness and Eddington ratio as found in the LLAGNs. We further perform the detailed modeling of the
  SED for 20 LLAGNs in our sample from the radio to X-ray waveband. We find that the SED of LLAGNs can be
  well described by our model, where the X-ray emission is produced predominantly by the inverse Compton scattering of
  the seed synchrotron photons produced by both thermal and nonthermal electrons in the RIAF, while the radio
  emission mainly come from the nonthermal electrons.

  It is now established that RQ AGNs are not radio-silent, and do emit the radio emission at some level. \citet{hu01} found
  that 85\% of the nearby Seyfert nuclei show the nuclear radio emission at 5 GHz, and their typical radio morphology is a compact core
  (either unresolved or slightly resolved). \citet{an04} observed six nearby LLAGNs with high-resolution VLBA, and found that
  the radio emission is still unresolved even at milliarcsecond scale, which roughly corresponds to several to ten thousands
  gravitational radii. Therefore, contrary to the RL AGNs which have been observed with
  jet or core-jet structures directly, the physical origin of the radio emission in RQ AGNs
  is much unclear. The most popular model is that there may exist a scaled-down version of the jet in RQ AGNs.
  The coupled RIAF-jet model has been explored to fit the multiwavelength spectrum of LLAGNs \citep[e.g.,][]{wu07,yu09,yyh11,ne12}.
  We note that it is reasonable to apply the RIAF-jet model to the sources with observed jet structures (e.g., FR Is),
  while it is still controversial whether the same model can be used to the LLAGNs
  with only compact radio cores, where these sources have no evident jet
  structure even at scale of several thousand gravitational radii. Another possibility is that the radio emission in
  RQ AGNs also originate in the hot plasma (e.g., RIAF in LLAGNs and disk-corona in bright AGNs), where the turbulence, weak shocks
  and/or magnetic reconnection events may occasionally accelerate a fraction of the electrons to a harder power-law tail.
  The synchrotron emission from these nonthermal power-law electrons may account for the radio emission from the compact cores
  of these LLAGNs. The hybrid thermal-nonthermal synchrotron emission from the RIAF have well reproduced the two components of
  the radio spectrum and also the X-ray flares observed in Sgr A* \citep[e.g.,][]{yu03}.
  We confirm the result that the pure RIAF always underpredicts the radio emission for nearby LLAGNs \citep[][]{wu05}.
  However, the low-frequency radio emission of the compact cores in LLAGNs can be accounted if only a small
  fraction of the viscous dissipation energy in the accretion flow goes into accelerating electrons
  to a nonthermal power-law distribution (see Figures \ref{f4}-\ref{f5}). It is well established that magnetic reconnection should
  be unavoidable and play an important role in converting the magnetic energy into the particles \citep[e.g.,][]{hb02,gu08}. Both the
  magnetic reconnection itself \citep[e.g.,][]{di10} or the diffusive shock caused by the violent plasma motions in the magnetic reconnection region
  \citep[e.g.,][]{sp88} can naturally accelerate a small fraction of the electrons to be the power-law distribution.
  A small fraction of the power-law electrons presented in the RIAF will not affect the global structure of
  the RIAF \citep[e.g.,][]{oz00}.
  Therefore, our work provide the possibility that the radio emission of nearby LLAGNs
  may originate in the nonthermal electrons of the hot accretion flow. It may be the similar case for other bright RQ AGNs, where the
  radio emission is dominated by the nonthermal electrons in the corona. \citet{lb08} found the similarities of the radio/hard X-ray
  correlation in RQ AGNs and magnetically active stars, which support that the radio emission of RQ AGNs also dominantly come from
  the corona as that of active stars. We will further test this issue in a subsequent work for bright RQ AGNs considering the possible
  physical mechanism for the production of thermal and nonthermal electrons in the corona above the SSD.

  \citet{ho02} found a strong anticorrelation between the radio loudness and Eddington ratio for the LLAGNs \citep[see also][]{si07,pa07,yo12}.
  We investigate the dependence of the mass-corrected radio loudness, $R_{\rm M}=R_{\rm X}/M_{\rm BH}^{0.4}$,
  on the Eddington ratio, $L_{2-10\rm keV}/L_{\rm Edd}$, where the scaling of $M_{\rm BH}^{0.4}$ is derived from our model as that
  $L_{\rm X}\propto M_{\rm BH}$ and $L_{\rm R}\propto M_{\rm BH}^{1.4}$. We find that the mass-corrected radio loudness
  is still anticorrelate with the Eddington ratio, where both quantities are not affected by the mass term.
  It is interesting to note that the dependence of the radio emission on the BH mass, $L_{\rm R}\propto M_{\rm BH}^{1.4}$,
  in our model is similar to that of jet model \citep[e.g.,][]{hs03} and that derived from the observed fundamental plane
  of BH activity \citep[][]{mhd03,fa04}. Therefore, it should be cautious to get the conclusion that the
  origin of the radio and X-ray emission is similar for the stellar-mass BH X-ray binaries (XRBs) and the supermassive BH AGNs based
  only on the fundamental plane equation of the BH activity, since that the dependence of the radio
  emission on the BH mass is similar for our model and the jet model.
  We find that the relation between the radio luminosity and X-ray luminosity is $L_{\rm R}\propto L_{\rm X}^{0.4}$ at
  a given BH mass in our model based on $L_{\rm R}\propto \dot{m}^{0.8}$ and $L_{\rm X}\propto \dot{m}^{2.0}$. Our model prediction
  on the radio$-$X-ray correlation is much shallower than that observed in XRBs \citep[e.g.,][]{ga03,co03} or that predicted
  by accretion-jet model \citep[e.g.,][]{yc05}, where $L_{\rm R}\propto L_{\rm X}^{0.7}$.
  Our results provide a diagnostic that can distinguish the possible origin of the radio emission in LLAGNs by
  investigating the radio$-$X-ray correlation for LLAGNs with the same or similar BH masses. However, our present sample
  is still limited, which prevent us to further test this issue.

  From a theoretical perspective, there are still some uncertainties in the current RIAF model with hybrid thermal and nonthermal electrons.
  The dependence of the spectrum on the model parameters have been explored in former works \citep[e.g.,][]{qn99,oz00}. The RIAF
  spectrum is not sensitive to some parameters (e.g., $\alpha$, $\beta$), we adopt the typical values $\alpha=0.3$ and $\beta=0.9$,
  which are constrained from observations and/or simulations. Wind parameter $s=0.3$ is adopted directly, which is constrained
  from  observations of Sgr A*. We adopt the radius $R_{\rm out}=5000 R_{\rm g}$ as the outer boundary.
  We find that the X-ray emission mainly come from the innermost region of the accretion flow, which is less affected by
  the outer boundary condition if it is larger than several tens gravitational radii. However, we find that the low-frequency radio emission come
  from the much larger regions from several tens to several thousand gravitational radii. The LLAGNs in our sample have very low
  Eddington ratios ($10^{-7} \lesssim L_{\rm 2-10\ keV}/L_{\rm Edd}\lesssim 10^{-4}$), and the whole accretion may be through RIAF.
  \citet{lt09} proposed that the truncation radius should be around one thousand to several tens of thousands gravitational
  radii (if the possible outer SSD is present) for $\dot{m}\sim10^{-4}-10^{-2}$ as adopted in fitting the
  SED of the LLAGNs in our sample. Therefore, the assumption of the outer boundary $R_{\rm out}=5000 R_{\rm g}$ should
  be reasonable, and will not affect our main results. The region of the radio emission from the nonthermal electrons in RIAF
  is more or less consistent with the size of the compact radio cores as observed by the VLBI for some nearby LLAGNs \citep[e.g.,][]{wh06}.
  We neglect the contribution of the emission from the possible outer SSD (if present) in our SED fittings,
  since that we still don't know whether the outer SSD still exists in the LLAGNs with extreme low Eddington ratios or not,
  and the optical emission also easily suffer the contamination from the host galaxies or contaminate by the nuclear starbursts.
  For example, \citet{sb05} reported an evidence of recent starburst formation in the nucleus of NGC 1097, and its optical/UV emission
  may dominated by the stellar contribution. We find that our model prediction on the optical emission is slightly smaller than the
  observed emission but higher than the lower limits of the possible intrinsic AGN optical emission that derived from optical variabilities.
  Therefore, the observed optical emission of LLAGNs is roughly consistent with our model predictions even without considering the possible
  contribution from the outer SSD (e.g., Figures \ref{f4}-\ref{f5}). The parameter
  $\delta$, describing the fraction of the turbulent dissipation that directly heats the electrons in the flow, is still unclear,
  which may be in the range of several percent up to several tens percent based on recent simulations \citep[e.g.,][]{sh07} and/or model fitting of the Sgr A*
   \citep[][]{yu03,yu06}. We find that most of LLAGNs in our sample can be better fitted if $\delta=0.1$ is adopted (see Figures
   \ref{f4}-\ref{f5}). It should be noted that the adopted $\delta=0.1$ is slightly smaller than that used in Sgr A* \citep[$\delta=0.3$,][]{yu03},
   which may caused by different microphysics since that the accretion rate in Sgr A* much smaller than that used in our
   work. Our results will be roughly unchanged even if we use $\delta=0.3$ as that used in Sgr A* \citep[e.g.,][]{yu06}, but
   the model fitting of the X-ray emission is less perfect as that with $\delta=0.1$.
  The values of $p$ and $\eta$ that describing the nonthermal electrons is not strongly constrained, but which are degenerate for low-frequency
  radio emission \citep[e.g.,][]{oz00}. We fix the typical value of $p$=3.0, and then adjust the parameter $\eta$ to fit the radio emission of LLAGNs
  in our sample. We find that only a small fraction of the energy in nonthermal electrons is sufficient to produce the radio emission at low frequencies.
  The value of $\eta$ can even be smaller if the parameter $p$ is also smaller. Our fitting results show that the value of $\eta$ in these LLAGNs of our
  sample is more or less consistent with that constrained for Sgr A* \citep[e.g., $\sim$1\% in][]{yu03,yu06} and that derived from the gamma-ray
  background \citep[e.g., $\sim$4\% in ][]{in08}.

\acknowledgments We appreciate Feng Yuan, Xinwu Cao and the anonymous referee for helpful comments and suggestions.
We also thank the members of HUST astrophysics group for useful discussions.
This work is supported by the National Basic Research Program of China (2009CB824800),  the NSFC (grants
11103003, 11133005 and 11173011), the Doctoral Program of Higher Education (20110142120037), the
Fundamental Research Funds for the Central Universities (HUST:
2011TS159).



\end{document}